\documentclass[11pt,a4paper]{article} \textwidth=15cm
\usepackage{latexsym}
\usepackage{amsfonts}
\usepackage{amssymb, amsfonts, amsmath}
\usepackage{amsbsy}
\date{}

\newtheorem{thm}{Theorem}[section]
 
 \newtheorem{lem}[thm]{Lemma}

  \newtheorem{defn}[thm]{Definition}
 \newtheorem{rem}[thm]{Remark}
 \numberwithin{equation}{section}

\begin{document}

\author{Mukhiddin I. Muminov and Tulkin H. Rasulov}
\title{\bf The Faddeev Equation and Essential Spectrum of a
Hamiltonian in Fock Space} \maketitle

\begin{abstract}
A Hamiltonian (model operator) $H$ associated to a quantum system
describing three particles in interaction, without conservation of
the number of particles, is considered. The Faddeev type system of
equations for eigenvectors of $H$ is constructed. The essential
spectrum of $H$ is described by the spectrum of the channel
operator.
\end{abstract}


\vspace{0.1cm}


\section{INTRODUCTION}

The essential spectrum of the systems with a fixed number of
particles has been studied in many articles, see for example, for
the continuous case \cite{RS-4,Zhis} and for the lattice case
\cite{LM,MM,Rab-Roch}.

In quantum field theory, condensed matter physics and the theory of
chemical reactions naturally occur the quantum systems with non
conserved number of particles. Often, the number of particles can be
arbitrary large as in cases involving photons (see e.g. \cite{BFS}),
in other cases, such as scattering of spin waves on defects,
scattering massive particles and chemical reactions, there are only
participants at any given time, though their number can be change.

Recall that the study of systems describing $N$ $(1 \leq N <
\infty)$ particles in interaction, without conservation of the
number of particles is reduced to the investigation of the spectral
properties of self-adjoint operators, acting in the {\it cut
subspace} ${\cal H}^{(N)}$ of Fock space, consisting of $n\leq N$
particles \cite{Frid,HSp,Min-Sp,Mog,SSZ,Zhu-Minl}.

The perturbation problem of an operator (the Friedrichs model), with
point and continuous spectrum (which acts in ${\cal H}^{(2)}$) has
played a considerable role in the study of spectral problems
connected to the quantum theory of fields \cite{Frid}.

A two level atom coupled to the radiation field was considered in
\cite{HSp} and using a Mourre type estimate, a complete spectral
characterization of the spin boson Hamiltonian was studied for
sufficiently small, but nonzero coupling. In \cite{SSZ} the quantum
systems with non conserved, but finite number of particles was
considered, and for such systems, geometric and commutator
techniques were developed, which were used to find the location of
the spectrum, to prove the absence of singular continuous spectrum
and identify accumulation points of the discrete spectrum.

In the present paper we consider the model operator $H$ associated
to a system describing three particles in interaction without
conservation of the number of particles, acting in ${\cal H}^{(3)},$
which is a lattice analog of the spin-boson Hamiltonian
\cite{HSp,Min-Sp}. Note that this operator, not studied before and
can be considered as a generalization of the above models studied in
\cite{ALR-1,ALR-2,LR-1,LR-2,R-1,R-2,R-3,Yod-Mum}.

The Faddeev equation and the location of the essential spectrum for
the similar to $H$ model operators acting in symmetric and non
symmetric Fock spaces have been studied in \cite{R-3,Yod-Mum} in the
case when the operators $V_i,\,i=1,2$ (defined below) are partial
integral operators generated kernels. But the techniques developed
in that papers are not applicable to the more general case,
considered in the present paper.

We obtain the following results:

(i) The Faddeev equation for the eigenvectors of $H$ is
constructed.

(ii) We describe the location of the essential spectrum of the
operator $H$ in terms of the spectrum of the channel operator
$\hat{H}.$

The paper is organized as follows. In Section 2, the model operator
$H$ is introduced and the main results are stated. In Section 3 the
spectrum of $\hat{H}$ is described by the spectrum of a family of
generalized Friedrichs models and some auxiliary statements, which
plays an important role in the proof of the main results of the
paper, are proven. In Section 4 we obtain an analogue of the Faddeev
type system of integral equations for the eigenfunctions of $H$
(Theorem \ref{About T(z)}) and prove that the essential spectrum of
$H$ coincides with the spectrum of the channel operator $\hat{H}$
(Theorem \ref{ess of H}). In section 5 is given an example of
calculation of the essential spectrum of $H,$ which shows the
efficiency of the proposed method of calculation of the essential
spectrum.

Throughout this paper we adopt the following convention: Denote by
${\bf T}^\nu$ the $\nu$-dimensional torus, the cube $(-\pi,\pi]^\nu$
with appropriately identified sides. The torus ${\bf T}^\nu$ will
always be considered as an abelian group with respect to the
addition and multiplication by real numbers regarded as operations
on the $\nu$-dimensional space ${\bf R}^\nu$ modulo $(2 \pi {\bf
Z})^\nu,$ where ${\bf Z}$ is the one-dimensional lattice.

\section{THE MODEL OPERATOR AND MAIN RESULTS}

Let us introduce some notations used in this work. Let ${\bf C}$ be
the field of complex numbers, $ L_2({\bf T}^\nu)$ be the Hilbert
space of square integrable (complex) functions defined on ${\bf
T}^\nu$ and $ L_2^{s}(({\bf T}^\nu)^2)$ be the Hilbert space of
square integrable (complex) symmetric functions defined on $({\bf
T}^\nu)^2.$

Denote by ${\cal H}$ the direct sum of spaces ${\cal H}_0={\bf C},$
${\cal H}_1=L_2({\bf T}^\nu)$ and ${\cal H}_2=L_2^{s}(({\bf
T}^\nu)^2),$ that is, ${\cal H}={\cal H}_0 \oplus {\cal H}_1 \oplus
{\cal H}_2.$

The Hilbert space ${\cal H}$ is called the {\it "three-particle
cut subspace"} of the Fock space.

Let $H_{ij}$ be annihilation (creation) operators \cite{Frid}
defined in the Fock space for $i<j$ ($i>j$). We note that in
physics, an annihilation operator is an operator that lowers the
number of particles in a given state by one, a creation operator is
an operator that increases the number of particles in a given state
by one, and it is the adjoint of the annihilation operator.

In this paper we consider the case, where the number of
annihilations and creations of the particles of the considering
system equal to 1. It means that $H_{ij}\equiv 0$ for all $|i-j|>1.$
So, a model operator $H$ associated to a system describing three
particles in interaction, without conservation of the number of
particles, acts in the Hilbert space ${\cal H}$ as a matrix operator
$$
H=\left( \begin{array}{cccc}
H_{00} & H_{01} & 0 \\
H_{10} & H_{11} & H_{12}\\
0 & H_{21} & H_{22} \\
\end{array}
\right).
$$

Let its components $H_{ij}: {\cal{H}}_j\to {\cal{H}}_i,\,\
i,j=0,1,2$ are defined by the rule
$$
(H_{00}f_0)_0=w_0f_0,\,\, (H_{01}f_1)_0=\int\limits_{{\bf T}^\nu}
v_0(s)f_1(s)ds,\,\,  (H_{10}f_0)_1(p)=v_0(p)f_0,
$$
$$
(H_{11}f_1)_1(p)=w_1(p)f_1(p),\quad
(H_{12}f_2)_1(p)=\int\limits_{{\bf T}^\nu} v_1(s)f_2(p,s)ds,
$$
$$
(H_{21}f_1)_2(p,q)=\frac{1}{2}(v_1(p)f_1(q)+v_1(q)f_1(p)),
$$
$$
H_{22}=H_{22}^0-V_{1}-V_{2},\quad
(H_{22}^0f_2)_2(p,q)=w_2(p,q)f_2(p,q),
$$
$$
(V_{1}f_2)_2(p,q)=\int\limits_{{\bf T}^\nu} v_2(p,s)f_2(s,q)ds,
(V_{2}f_2)_2(p,q)=\int\limits_{{\bf T}^\nu} v_2(s,q)f_2(p,s)ds.
$$

Here $f_i\in {\cal H}_i,\,i=0,1,2,$ $w_0$ is a real number,
$w_1(\cdot),\,v_{i}(\cdot),\,i=0,1$ are real-valued continuous
functions on ${\bf T}^\nu,$ $w_2(\cdot,\cdot)$ and
$v_2(\cdot,\cdot)$ are real-valued continuous symmetric functions
on $({\bf T}^\nu)^2.$

Under these assumptions the operator $H$ is bounded and self-adjoint
in ${\cal H}.$

Set
$$
\overline{{\cal H}}_0 = {\cal H}_0,\quad \overline{{\cal H}}_1 =
{\cal H}_1,\quad \overline{{\cal H}}_2=L_2(({\bf T}^\nu)^2) \quad
\mbox{and} \quad {\cal H}^{(n,m)}=\bigoplus_{i=n}^m \overline{{\cal
H}}_i,\quad 0\leq n<m \leq 2,
$$
where $ L_2(({\bf T}^\nu)^2)$ is the Hilbert space of square
integrable (complex) functions on $({\bf T}^\nu)^2.$

Throughout the paper we additionally assume that the operators
$V_{i},\,i=1,2$ acting in the Hilbert space $\overline{{\cal H}}_2$
are positive and use this fact without comments. Denote by
$\tilde{V}_i,\,i=1,2$ a positive square root of the operators
$V_{i},\,i=1,2.$ Then the operators $\tilde{V}_i,\,i=1,2$ has form
(see Lemma \ref{V-1-positive})
\begin{equation}\label{roots of V_i}
(\tilde{V}_1f_2)(p,q)=\int\limits_{{\bf T}^\nu}
\tilde{v}_2(p,s)f_2(s,q)ds,\,\,
(\tilde{V}_2f_2)(p,q)=\int\limits_{{\bf T}^\nu}
\tilde{v}_2(q,s)f_2(p,s)ds,\,\, f_2\in \overline{{\cal H}}_2.
\end{equation}

Here the kernel of $\tilde{V}_i,\,i=1,2$ formally denoted by
$\tilde{v}_2(\cdot,\cdot).$

To formulate our main results we introduce the channel operator
$\hat{H}$ acting in ${\cal{H}}^{(1,2)}$ by the following rule
$$
\hat{H}=\left( \begin{array}{cc}
H_{11} & \frac{1}{\sqrt{2}}\, H_{12}\\
\frac{1}{\sqrt{2}}\, H_{21}^{(1)} & H_{22}^0-V_2
\end{array}
\right)
$$
with
$$
(H_{21}^{(1)}f_1)(p,q)=v_1(q)f_1(p),\,f_1\in \overline{{\cal H}}_1.
$$

It is easy to show that the operator $\hat{H}$ is bounded and
self-adjoint in ${\cal{H}}^{(1,2)}.$

Let
$$
m = \min_{p,q\in {\bf T}^\nu} w_2(p,q),\quad M= \max_{p,q\in {\bf
T}^\nu} w_2(p,q).
$$

For each $z\in {\bf C}\setminus [m; M]$ we define the operator
matrices $A(z)$ and  $K(z)$ act in the Hilbert space ${\cal
H}^{(0,2)}$ as
$$
A(z)=\left( \begin{array}{ccc}
A_{00}(z) & 0 & 0\\
0 & A_{11}(z) & A_{12}(z)\\
0 & A_{21}(z) & A_{22}(z)
\end{array}
\right),\quad K(z)=\left( \begin{array}{ccc}
K_{00}(z) & K_{01}(z) & 0\\
K_{10}(z) & K_{11}(z) & K_{12}(z)\\
0 & K_{21}(z) & K_{22}(z)
\end{array}
\right),
$$
where the operators $A_{ij}(z): \overline{\cal H}_j \to
\overline{\cal H}_i ,\,\, i,j=0,1,2$ are defined as
$$
(A_{00}(z)g_0)_0=g_0,\quad (A_{11}(z)g_1)_1(p)=\Bigl(
w_1(p)-z-\frac{1}{2}\int\limits_{{\bf T}^\nu}
\frac{v_1^2(s)ds}{w_2(p,s)-z} \Bigr) g_1(p),
$$
$$
(A_{12}(z) g_2)_1(p)= \int\limits_{{\bf T}^\nu}
\frac{v_1(s)}{w_2(p,s)-z}\int\limits_{{\bf T}^\nu} \tilde{v}_2(t,s)
g_2(p,t)dtds,
$$
$$
(A_{21}(z)g_1)_2(p,q)=\frac{1}{2}\int\limits_{{\bf T}^\nu}
\frac{\tilde{v}_2(s,q)v_1(s)}{w_2(p,s)-z}ds\, g_1(p),
$$
$$
(A_{22}(z)g_2)(p,q)=g_2(p,q)-(\tilde{V}_2 R_{22}^0(z) \tilde{V}_2
g_2)(p,q)
$$
and the operators $K_{ij}(z): \overline{\cal{H}}_j \to
\overline{\cal{H}}_i ,\,\, i,j=0,1,2$ are defined as
$$
(K_{00}(z)g_0)_0=(w_0-z+1)g_0, \, K_{01}(z)\equiv
H_{01},\,K_{10}(z)\equiv -H_{10},
$$
$$
(K_{11}(z)g_1)_1(p)=\frac{v_1(p)}{2}\int\limits_{{\bf T}^\nu}
\frac{v_1(s)g_1(s)ds}{w_2(p,s)-z},
$$
$$
(K_{12}(z)g_2)_1(p)= \int\limits_{{\bf T}^\nu}
\frac{v_1(s)}{w_2(p,s)-z}\int\limits_{{\bf T}^\nu}
\tilde{v}_2(p,t)g_2(t,s)dtds,
$$
$$
(K_{21}(z)g_1)_2(p,q)=-\frac{v_1(p)}{2}\int\limits_{{\bf T}^\nu}
\frac{\tilde{v}_2(s,q)g_1(s)}{w_2(p,s)-z}ds,
$$
$$
(K_{22}(z)g_2)_2(p,q)=(\tilde{V}_2 R_{22}^0(z) \tilde{V}_1
g_2)(p,q),
$$
where $g_i \in \overline{{\cal H}}_i,\,i=0,1,2$ and
$R_{22}^0(z)=(H_{22}^0-z)^{-1}$ is the resolvent of the operator
$H_{22}^0.$

We note that for each $z\in {\bf C}\setminus [m; M]$ the operators
$K_{ij}(z),\, i,j=0,1,2$ belong to the Hilbert-Schmidt class and
therefore $K(z)$ is a compact operator.

Let $\sigma(\hat{H})$ be the spectrum of $\hat{H}.$ Since for any
fixed $z\in {\bf C}\setminus \sigma(\hat{H})$ the operator $A(z)$ is
bounded and invertible (see Lemma \ref{A(z) invertible}), for such
$z$ we can define the operator $T(z)=A^{-1}(z)K(z).$

Now we give the main results of the paper.

The following theorem establishes a connection between eigenvalues
of $H$ and $T(z).$
\begin{thm}\label{About T(z)} The number $z\in {\bf C}\setminus
\sigma(\hat{H})$ is an eigenvalue of the operator $H$ if and only
if the number $\lambda=1$ is an eigenvalue of the operator $T(z).$
\end{thm}
\begin{rem} We point out that the equation $T(z)g=g$ is an analogue of
the Faddeev type system of integral equations for eigenvectors of
the operator $H$ and its played crucial role in our analysis of the
spectrum of $H.$
\end{rem}

The following theorem  describes the essential spectrum of the
operator $H.$

\begin{thm}\label{ess of H} The essential spectrum
of $H$ coincides with the spectrum of $\hat{H}.$
\end{thm}

Since the channel operator $\hat{H}$ has a more simple structure
than $H,$ Theorem \ref{ess of H} plays an important role in the next
investigations of the spectrum of $H.$ We note that by Lemma
\ref{spektr of channel} (see Section 4) Theorem \ref{ess of H}
describes the location of the essential spectrum of $H$ in terms of
the spectrum of the channel operator $\hat{H},$ where separated
two-particle and three-particle branches of this spectrum.

\section{SOME AUXILIARY STATEMENTS}

In this section we describe the spectrum of the channel operator
$\hat{H}.$ Using the decomposition into direct operator integrals
(see \cite{RS-4}) we reduce to study the spectral properties of the
operator $\hat{H}$ to the investigation of the spectral properties
of the family of operators $h(p),\, p\in {\bf T}^\nu$ defined below.
We also give some auxiliary statements that allow us to prove the
main results of the paper.

Let the operator $v$ act in ${\cal H}_1$ as
\begin{equation*}\label{operator v}
(vf)(p)=\int\limits_{{\bf T}^\nu} v_2(p,s)f(s)ds,\,f\in {\cal
H}_1.
\end{equation*}

\begin{lem}\label{about root of v}
The operator $v$ is positive and its positive square root
$\tilde{v}\equiv v^{\frac{1}{2}}$ has form
\begin{equation}\label{formula root of v}
(\tilde{v}f)(q)=\int\limits_{{\bf T}^\nu}
\tilde{v}_2(q,s)f(s)ds,\,f\in {\cal H}_1.
\end{equation}

Moreover, the function $\tilde{v}_2(\cdot,\cdot)$ is a square
integrable on $({\bf T}^\nu)^2$.
\end{lem}
{\bf Proof.} Since $v_2(\cdot,\cdot)$ is a continuous function on
$({\bf  T}^\nu)^2$ we have
$$
\int\limits_{{\bf T}^\nu} |v_2(s,s)|ds<\infty.
$$
The function $v(\cdot, \cdot)$ is symmetric and hence the last
inequality means that the operator $v$ belongs to the trace class.
From the positivity of $V_i,\,i=1,2$ it follows that the operator
$v$ is also positive. Therefore, every nontrivial eigenvalue
$\lambda_k$ of $v$ are positive. By the Hilbert-Schmidt theorem we
have
$$
v=\sum\limits_k \lambda_k(\varphi_k,\cdot)\varphi_k
$$
with $\sum\limits_k \lambda_k< \infty,$ where $\varphi_k$ is the
eigenfunction of the operator $v$  corresponding to the eigenvalue
$\lambda_k.$ Then
 $$
 \tilde{v}=\sum_k \sqrt{\lambda_k}(\varphi_k,\cdot)\varphi_k.
 $$
Taking into account $\sum\limits_k \lambda_k< \infty$ we obtain that
$\tilde{v}$ is  the Hilbert-Schmidt operator. Therefore the kernel
$\tilde{v}_2(\cdot,\cdot)$ of the integral operator $\tilde{v}$ is a
square integrable function. $\Box$

Let $I_i,\,i=0,1,2$ be an identity operator in ${\cal
H}_i,\,i=0,1,2.$

\begin{lem}\label{V-1-positive} The positive square root of $V_i,\,i=1,2$ has form
(\ref{roots of V_i}).
\end{lem}
{\bf Proof.} The operators $V_i, i=1,2$ can be decomposed as
$$
V_1= v\otimes I_1,\quad V_2=I_1\otimes v.
$$

By Lemma \ref{about root of v} the operator $v$ is positive and its
positive square root has form (\ref{formula root of v}). Now it is
easy to check that $\tilde{V}_1= \tilde{v}\otimes I_1$ and
$\tilde{V}_2=I_1\otimes \tilde{v}.$ $\Box$

We now study the operator $\hat{H},$ which commutes with any
multiplication operator $U_\alpha$ by the bounded function
$\alpha(\cdot)$ on ${\bf T}^\nu:$
$$
U_\alpha \left( \begin{array}{cc}
g_1(p)\\
g_2(p,q)
\end{array}
\right)=\left( \begin{array}{cc}
\alpha(p) g_1(p)\\
\alpha(p) g_2(p,q)
\end{array}
\right),\, \left( \begin{array}{cc}
g_1\\
g_2
\end{array}
\right) \in {\cal{H}}^{(1,2)}.
$$

Therefore the decomposition of the space ${\cal H}^{(1,2)}$ into the
direct integral (see XIII.16 in \cite{RS-4})
$$
{\cal H}^{(1,2)}= \int\limits_{{\bf T}^\nu} \oplus \,{\cal
H}^{(0,1)} dp
$$
yields the decomposition into the direct integral
\begin{equation}\label{decom hat H}
\hat{H}= \int\limits_{{\bf T}^\nu} \oplus \,h(p)dp,
\end{equation}
where the family of the bounded and self-adjoint operators
$h(p),\,p\in {\bf T}^\nu$ acts in ${\cal H}^{(0,1)}$ as
\begin{equation}\label{formula of h(p)}
h(p)=\left( \begin{array}{cc}
h_{00}(p) & h_{01}\\
h_{10} & h_{11}^0(p)-v
\end{array}
\right)
\end{equation}
with the entries
$$
(h_{00}(p)f_0)_0=w_1(p)f_0,\,\, (h_{01}f_1)_0= \frac{1}{\sqrt{2}}
\int\limits_{{\bf T}^\nu} v_1(s)f_1(s)ds,
$$
$$
(h_{10}f_0)_1(q)= \frac{1}{\sqrt{2}} v_1(q)f_0,\quad
(h_{11}^{0}(p)f_1)_1(q)=w_2(p,q)f_1(q).
$$

Let the operator $h_0(p),\,p\in {\bf T}^\nu$ act in ${\cal
H}^{(0,1)}$ as
\begin{equation*}
h_0(p)=\left( \begin{array}{cc}
0 & 0\\
0 & h_{11}^{0}(p)\\
\end{array}
\right),\,p\in {\bf T}^\nu.
\end{equation*}

The perturbation $h(p)-h_0(p),\,p\in {\bf T}^\nu$ of the operator
$h_0(p),\,p\in {\bf T}^\nu$ is a compact operator. Therefore in
accordance with the invariance of the essential spectrum under
compact perturbations the essential spectrum $\sigma_{ess}(h(p))$ of
$h(p),\,p\in {\bf T}^\nu$ fills the following interval on the real
axis:
\begin{equation*}
\sigma_{ess}(h(p))=[m(p); M(p)],
\end{equation*}
where the numbers  $m(p)$ and $M(p)$ are defined by
\begin{equation*}
m(p)= \min_{q\in {\bf T}^\nu} w_2(p,q),\quad M(p)= \max_{q\in {\bf
T}^\nu} w_2(p,q).
\end{equation*}

For any fixed $p\in {\bf T}^\nu$ we define the matrix operator
$A_0(p;z)$ act in ${\cal H}^{(0,1)}$ as
\begin{equation}\label{A_0}
A_0(p;z)=\left( \begin{array}{cc}
A_{00}(p;z) & A_{01}(p;z)\\
A_{10}(p;z) & A_{11}(p;z)
\end{array}
\right),\,z\in {\bf C} \setminus \sigma_{ess} (h(p)),
\end{equation}
where
$$
(A_{00}(p;z)g_0)_0=\Bigl( w_1(p)-z-\frac{1}{2}\int\limits_{{\bf
T}^\nu} \frac{v_1^2(s)ds}{w_2(p,s)-z}-1 \Bigr) g_0,
$$
$$
(A_{01}(p;z) g_1)_0= \int\limits_{{\bf T}^\nu}
\frac{v_1(s)}{w_2(p,s)-z}\int\limits_{{\bf T}^\nu} \tilde{v}_2(t,s)
g_1(t)dtds,
$$
$$
(A_{10}(p;z)g_0)_1(q)=\frac{1}{2}\int\limits_{{\bf T}^\nu}
\frac{\tilde{v}_2(s,q)v_1(s)}{w_2(p,s)-z}ds\, g_0,
$$
$$
(A_{11}(p;z)g_1)_1(q)=-\int\limits_{{\bf
T}^\nu}\frac{\tilde{v}_2(q,t)}{w_2(p,t)-z}\int\limits_{{\bf T}^\nu}
\tilde{v}_2(t,s)g_1(s)dsdt.
$$

We note that for any fixed $p \in {\bf T}^\nu$ and $z \in {\bf C}
\setminus \sigma_{ess} (h(p))$  the operator $A_0(p\,; z)$ belongs
to the trace class. Therefore (see \cite{RS-4}) the determinant
$det[E+ A_0(p\,; z)]$ of the operator  $E+ A_0(p\,; z)$ is well
defined, where $E=diag \{I_0, I_1\}.$

The following lemma establishes a connection between the
eigenvalues of $h(p),\,p\in {\bf T}^\nu$ and the zeroes of the
function $ det[E+ A_0(p\,; \cdot)],\,p\in {\bf T}^\nu.$

\begin{lem}\label{det A_0} The number $z \in {\bf
C} \setminus \sigma_{ess} (h(p))$ is an eigenvalue of the operator
$h(p),\,p \in {\bf T}^\nu$ if and only if $det[E+ A_0(p\,; z)]=0.$
\end{lem}

{\bf Proof.} Let the number $z \in {\bf C}\setminus \sigma_{ess}
(h(p))$ be an eigenvalue of the operator $h(p),\,p \in {\bf T}^\nu$
and $f=(f_0,f_1)\in {\cal H}^{(0,1)}$ be the corresponding
eigenvector, i.e. the equation $h(p)f=zf$ or the system of equations
$$
(w_1(p)-z)f_0+ \frac{1}{\sqrt{2}} \int\limits_{{\bf T}^\nu}
v_1(s)f_1(s)ds=0;
$$
\begin{equation}\label{sistem h_3}
 \frac{1}{\sqrt{2}} v_1(q)f_0+(w_2(p,q)-z)f_1(q)-\int\limits_{{\bf T}^\nu}v_2(q,s)f_1(s)ds=0
\end{equation}
has a nontrivial solution $f=(f_0,f_1)\in {\cal H}^{(0,1)}.$

Since $z \in {\bf C}\setminus \sigma_{ess} (h(p))$ from the second
equation of system (\ref{sistem h_3}) we  find
\begin{equation}\label{f_1}
f_1(q)=\frac{(\tilde{v} \tilde{f}_1)(q)}{w_2(p,q)-z}-
\frac{1}{\sqrt{2}} \frac{v_1(q) f_0}{w_2(p,q)-z},
\end{equation}
where the operator $\tilde{v}$ is defined by (\ref{formula root of
v}) and
\begin{equation}\label{tilde f_1}
\tilde{f}_1(q)=(\tilde{v} f_1)(q).
\end{equation}

Substituting the expression (\ref{f_1}) for $f_1$ into the first
equation of system (\ref{sistem h_3}) and the equality (\ref{tilde
f_1}), we get that the system of equations (\ref{sistem h_3}) has
nontrivial solution if and only if the system of equations
$$
\Bigl( w_1(p)-z-\frac{1}{2}\int\limits_{{\bf T}^\nu}
\frac{v_1^2(s)ds}{w_2(p,s)-z} \Bigr) f_0+ \int\limits_{{\bf T}^\nu}
\frac{v_1(s)}{w_2(p,s)-z}\int\limits_{{\bf T}^\nu} \tilde{v}_2(t,s)
\tilde{f}_1(t)dtds=0;
$$
$$
\frac{1}{\sqrt{2}}\int\limits_{{\bf T}^\nu}
\frac{\tilde{v}_2(s,q)v_1(s)}{w_2(p,s)-z}ds\, f_0
+\tilde{f}_1(q)-\int\limits_{{\bf
T}^\nu}\frac{\tilde{v}_2(q,t)}{w_2(p,t)-z}\int\limits_{{\bf T}^\nu}
\tilde{v}_2(t,s)\tilde{f}_1(s)dsdt=0
$$
or the equation
$$
E\Phi+A_0(p;z)\Phi =0,\quad \Phi=(f_0, \tilde{f}_1)\in {\cal
H}^{(0,1)}
$$
has a nontrivial solution, i.e. $det [E+ A_0(p\,; z)]=0.$ $\Box$

By Lemma \ref{det A_0} the number $z$ belongs to the discrete
spectrum of $h(p)$ if and only if $det [E+ A_0(p\,; z)]=0.$ It
immediately follows the following equality
\begin{equation}\label{disc spec h(p)}
\sigma_{disc}(h(p))=\{z\in {\bf C}\setminus \sigma_{ess}(h(p)): det
[E+ A_0(p\,; z)]=0\},\,p\in {\bf T}^\nu.
\end{equation}

\begin{lem}\label{spektr of channel} For the spectrum $\sigma(\hat{H})$ of
$\hat{H}$ the equality
$$
\sigma(\hat{H})=\bigcup\limits_{p\in {\bf T}^\nu}
\sigma_{disc}(h(p)) \cup [m; M]
$$
holds.
\end{lem}

{\bf Proof.} The assertion of this lemma follows from the
representation (\ref{decom hat H}), the equalities
$$
\sigma(h(p))=[m(p); M(p)] \cup \sigma_{disc}(h(p)), \quad
\bigcup\limits_{p\in {\bf T}^\nu} [m(p); M(p)]=[m; M]
$$
and the theorem on the spectrum of decomposable operators (see
\cite{RS-4}). $\Box$

Now we introduce the new subsets of the essential spectrum of $H.$
\begin{defn}\label{branches} The sets $\sigma_{two}(H)=
\bigcup\limits_{p\in {\bf T}^\nu} \sigma_{disc}(h(p))$ and
$\sigma_{three}(H)=[m; M]$ are called two-particle and
three-particle branches of the essential spectrum of $H,$
respectively.
\end{defn}

\begin{lem}\label{A(z) invertible} The operator $A(z),\,z\in {\bf C}\setminus
\sigma_{three}(H)$ is bounded and
invertible if and only if $z\in {\bf C}\setminus \sigma(\hat{H}).$
\end{lem}

{\bf Proof.} Let us introduce the operator matrix $A_0(z)$ acting in
${\cal H}^{(1,2)}$ as
$$
A_0(z)=\left( \begin{array}{cc}
A_{11}(z) & A_{12}(z)\\
A_{21}(z) & A_{22}(z)
\end{array}
\right).
$$
By the definition of $A(z)$ and $A_0(z)$ we have that the operator
$A(z)$ is invertible if and only if the operator $A_0(z)$ is
invertible.

In analogy with the operator $\hat{H}$ one can give the
decomposition
\begin{equation}\label{decomp A}
A_0(z)= \int\limits_{{\bf T}^\nu} \oplus [E+A_0(p\,; z)]dp,
\end{equation}
where the operator $A_0(p\,; z)$ is defined by
(\ref{A_0}).

By Lemmas \ref{det A_0} and \ref{spektr of channel} for any fixed
$p\in {\bf T}^\nu$ and $z\in {\bf C} \setminus \sigma(\hat{H})$ we
have $det[E+ A_0(p;z)]\neq 0.$ Therefore, for any fixed $z\in {\bf
C} \setminus \sigma(\hat{H})$ the operator $A_0(z)$ is invertible.
Conversely trivially follows from the decomposition (\ref{decomp
A}).
 $\Box$

\section{PROOF OF THE MAIN RESULTS}

In this section we prove Theorems \ref{About T(z)} and \ref{ess of
H}.

{\bf Proof of Theorem \ref{About T(z)}.} Let $ z\in {\bf C}
\setminus \sigma(\hat{H})$ be an eigenvalue of the operator $H$ and
$f=(f_0,f_1,f_2)\in {\cal H}$ be the corresponding eigenvector, that
is, the equation $Hf=zf$ or the system of equations
$$
((H_{00}-zI_0)f_0)_0+ (H_{01}f_1)_0=0;
$$
\begin{equation}\label{sistem equations}
(H_{10}f_0)_1(p)+((H_{11}-zI_1)f_1)_1(p)+(H_{12}f_2)_1(p)=0;
\end{equation}
$$
(H_{21}f_1)_2(p,q)+((H_{22}-z I_2)f_2)_2(p,q)=0
$$
have a nontrivial solution $f=(f_0,f_1,f_2)\in {\cal H}.$ Since
$z\not\in \sigma_{three}(H),$ from the third equation of the
system (\ref{sistem equations}) for $f_2,$ we have
\begin{equation}\label{ffff}
f_2(p,q)=(R_{22}^0(z) V_1f_2)(p,q)+
(R_{22}^0(z)V_2f_2)(p,q)-(R_{22}^0(z) H_{21}f_1)(p,q).
\end{equation}
 Let
\begin{equation}\label{tilde f}
\widetilde{f}_2(p,q)=(\tilde{V}_2f_2)(p,q).
\end{equation}
Then
$$
(\tilde{V}_2 \widetilde{f}_2)(p,q)=(\tilde{V}_1
\widetilde{f}_2)(q,p).
$$
 Therefore the  equality (\ref{ffff}) has form
\begin{equation}\label{f_2}
f_2(p,q)=(R_{22}^0(z)\tilde{V}_1 \widetilde{f}_2)(p,q)+ (R_{22}^0(z)
\tilde{V}_2\widetilde{f}_2)(p,q)-(R_{22}^0(z) H_{21}f_1)(p,q).
\end{equation}

Substituting the expression (\ref{f_2}) for $f_2$ into the system of
equations (\ref{sistem equations}) and the equality (\ref{tilde f})
we obtain that the system of equations
$$
f_0=(w_0-z+1)f_0 + \int\limits_{{\bf T}^\nu} v_0(s)f_1(s) ds;
$$
$$
\Bigl( w_1(p)-z-\frac{1}{2}\int\limits_{{\bf T}^\nu}
\frac{v_1^2(s)ds}{w_2(p,s)-z} \Bigr) f_1(p)+\int\limits_{{\bf
T}^\nu} \frac{v_1(s)}{w_2(p,s)-z}\int
\tilde{v}_2(t,s)\widetilde{f}_2(p,t)dtds=
$$
\begin{equation}\label{system-1}
-v_0(p)f_0+\int\limits_{{\bf T}^\nu} \frac{v_1(s)}{w_2(p,s)-z}\int
\tilde{v}_2(p,t)\widetilde{f}_2(t,s)dtds+
\frac{v_1(p)}{2}\int\limits_{{\bf T}^\nu}
\frac{v_1(s)f_1(s)ds}{w_2(p,s)-z};
\end{equation}
$$
\frac{1}{2}\int \frac{\tilde{v}_2(s,q)v_1(s)}{w_2(p,s)-z}ds f_1(p)+
\widetilde{f}_2(p,q)-(\tilde{V}_2R_{22}^0(z) \tilde{V}_2
\widetilde{f}_2)(p,q)=
$$
$$
(\tilde{V}_2R_{22}^0(z)\tilde{V}_1
\widetilde{f}_2)(p,q)-\frac{v_1(p)}{2}\int
\frac{\tilde{v}_2(s,q)f_1(s)}{w_2(p,s)-z}ds
$$
have a nontrivial solution if and only if the system of equations
(\ref{sistem equations}) has a nontrivial solution.

The system of equations (\ref{system-1}) can be written in the
following form
$$
A(z)\Phi=K(z)\Phi, \quad \Phi=(f_0,f_1,\widetilde{f}_2)\in {\cal
H}^{(0,2)}.
$$

By Lemma \ref{A(z) invertible} for each $z\in {\bf C} \setminus
\sigma(\hat{H})$ the operator $A(z)$ is invertible and hence the
following equation $\Phi=A^{-1}(z)K(z)\Phi $ or $ \Phi=T(z)\Phi$
has a nontrivial solution if and only if the system of equations
(\ref{system-1}) has a nontrivial solution. $\Box$

Now applying the Weyl criterion and Theorem \ref{About T(z)} we
prove Theorem \ref{ess of H}.

{\bf Proof of Theorem \ref{ess of H}.} The inclusion
$\sigma_{three}(H) \subset \sigma_{ess}(H)$ can be proven quite
similarly to the corresponding inclusion of \cite{LR-1}.

We prove that $\sigma_{two}(H) \subset \sigma_{ess}(H).$ Let $z_0\in
\sigma_{two}(H)$ be an arbitrary point.

Two cases are possible:

$z_0\in \sigma_{three}(H)$ or $z_0\not \in \sigma_{three}(H).$

 If $z_0\in \sigma_{three}(H),$
then $z_0\in \sigma_{ess}(H).$ Let $z_0\not \in \sigma_{three}(H),$
but $z_0\in \sigma_{two}(H).$ Then by Lemma \ref{A(z) invertible}
the operator $A(z_0)$ isn't invertible. It means that there exists
orthonormal system $\Phi^{(n)}=(0, f_1^{(n)},
\widetilde{f}_2^{(n)})$ such that $||A(z_0)\Phi^{(n)}||_{{\cal
H}^{(0,2)}}\to 0$ as $n\to +\infty.$

We choose a sequence of orthogonal vector-functions $\{f^{(n)}\}$ as
\begin{equation*}
f^{(n)} = \left ( \begin{array}{ccc}
0 \\
f_1^{(n)}(p)\\
f_2^{(n)}(p, q)
\end{array}
\right) ,
\end{equation*}
where
\begin{equation*}
f_2^{(n)}(p,q)=(R_{22}^0(z_0) \tilde{V}_1
\widetilde{f}^{(n)}_2)(p,q)+ (R_{22}^0(z_0)
\tilde{V}_2\widetilde{f}^{(n)}_2)(p,q)-(R_{22}^0(z_0)
H_{21}f_1^{(n)})(p,q).
\end{equation*}

We consider $(H-z_0)f^{(n)}$ and estimate its norm as
$$
||(H-z_0)f^{(n)}||_{{\cal H}}^2=||(H-z_0)f^{(n)}+
\widetilde{f}_2^{(n)}-\tilde{V}_2f_2^{(n)}
-(\widetilde{f}_2^{(n)}-\tilde{V}_2f_2^{(n)})||_{{\cal H}}\leq
$$
$$
||(A(z_0)-K(z_0))\Phi^{(n)}||_{{\cal
H}^{(0,2)}}^2+||\widetilde{f}_2^{(n)}-\tilde{V}_2f_2^{(n)}||_{\overline{{\cal
H}}_2}^2.
$$

Let
\begin{equation*}
(A(z_0)-K(z_0))\Phi^{(n)}= \left ( \begin{array}{ccc}
((A(z_0)-K(z_0))\Phi^{(n)})_0 \\
((A(z_0)-K(z_0))\Phi^{(n)})_1\\
((A(z_0)-K(z_0))\Phi^{(n)})_2
\end{array}
\right).
\end{equation*}

Since the operator $K(z_0)$ is a compact, we have
$||K(z_0)\Phi^{(n)}||_{{\cal H}^{(0,2)}}\to 0$ as $n\to +\infty.$
Therefore, from $||A(z_0)\Phi^{(n)}||_{{\cal H}^{(0,2)}}\to 0$  as
$n\to +\infty$ it follows that
\begin{equation}\label{estimate}
||(A(z_0)-K(z_0))\Phi^{(n)}||_{{\cal
H}^{(0,2)}}^2=||((A(z_0)-K(z_0))\Phi^{(n)})_0||_{\overline{{\cal
H}}_0}^2+ ||((A(z_0)-K(z_0))\Phi^{(n)})_1||_{\overline{{\cal
H}}_1}^2+
$$
$$
||((A(z_0)-K(z_0))\Phi^{(n)})_2||_{\overline{{\cal H}}_2}^2\leq
||A(z_0)\Phi^{(n)}||_{{\cal
H}^{(0,2)}}^2+||K(z_0)\Phi^{(n)}||_{{\cal H}^{(0,2)}}^2 \to 0
\end{equation}
 as $n\to +\infty.$ It follows that $||((A(z_0)-K(z_0))\Phi^{(n)})_i||_{\overline{{\cal
H}}_i}\to 0,\,i=0,1,2$ as $n\to \infty.$ Therefore from the equality
$$
||((A(z_0)-K(z_0))\Phi^{(n)})_2||_{\overline{{\cal
H}}_2}=||\widetilde{f}_2^{(n)}-\tilde{V}_2f_2^{(n)}||_{\overline{{\cal
H}}_2}
$$
and the relation (\ref{estimate}) we have that
$||(H-z_0)f^{(n)}||_{{\cal H}}\to 0$ as $n\to +\infty.$ This implies
that $z_0\in \sigma_{ess}(H).$ Since the point $z_0\in
\sigma_{two}({\hat H})$ is arbitrary, it follows that
$\sigma_{two}(H) \subset \sigma_{ess}(H).$

Now we prove the inclusion $\sigma_{ess}(H) \subset
\sigma(\hat{H}).$ Since for each $z\in {\bf C}\setminus
\sigma(\hat{H})$ the operator $K(z)$ is a compact and $A^{-1}(z)$ is
bounded, we have that $f(z)=A^{-1}(z)K(z)$ is a compact-valued
analytic function in ${\bf C}\backslash \sigma(\hat{H}).$ From the
self-adjointness of $H$ and Theorem \ref{About T(z)} it follows that
the operator $({\bf I}-f(z))^{-1}$ exists for all $Im z\not=0,$
where ${\bf I}$ is an identical operator in ${\cal H}^{(0,2)}.$ In
accordance with the analytic Fredholm theorem, we conclude that the
set
$$
\sigma(H)\setminus \sigma(\hat{H})=\{z: det({\bf I}-f(z))=0\}
$$
is discrete. Thus $\sigma(H)\setminus \sigma(\hat{H}) \subset
\sigma_{disc}(H)=\sigma(H)\setminus \sigma_{ess}(H).$ Therefore
the inclusion $\sigma_{ess}(H)\subset \sigma(\hat{H})$ holds.
$\Box$

\section{EXAMPLE}

In this section we consider the case $\nu=3$ and calculate the
essential spectrum of the operator $H$ in the case, where $w_0$ is
an arbitrary real number, $w_1(\cdot),\,v_{i}(\cdot),\,i=0,1$ are an
arbitrary real-valued continuous functions on ${\bf T}^3,$
$w_2(\cdot,\cdot)$ is an arbitrary real-valued continuous symmetric
function on $({\bf T}^3)^2$ and the function $v_2(\cdot,\cdot)$ has
form
\begin{equation}\label{formula for v_2}
v_2(p,q)=\sum_{i=1}^3 \cos (p_i - q_i),
\,p=(p_1,p_2,p_3),\,q=(q_1,q_2,q_3)\in {\bf T}^3.
\end{equation}

In this case for the kernel $\tilde{v}_2(\cdot, \cdot)$ of the
integral operator $\tilde{v}$ defined by (\ref{about root of v}) the
equality $\tilde{v}_2(p, q)=(4 \pi^3)^{-1} {v}_2(p, q)$ holds.

Additionally we also assume that the function $v_1(\cdot)$ is even
on ${\bf T}^3$ and the function $w_2(\cdot,\cdot)$ is even of any
coordinates on ${\bf T}^1$, for example,
$$
v_1(p)= \sum_{i=1}^3 \cos p_i,\, p=(p_1,p_2,p_3) \in {\bf T}^3;
$$
\begin{equation}\label{example for w_2}
w_2(p,q)= \sum_{i=1}^3 (2-\cos p_i - \cos q_i),\,
p=(p_1,p_2,p_3),\,q=(q_1,q_2,q_3)\in {\bf T}^3.
\end{equation}

From Theorem \ref{ess of H} and Lemma \ref{spektr of channel} it
follows that in the study the essential spectrum of $H$ the crucial
role plays the discrete spectrum of $h(p)$ defined by (\ref{formula
of h(p)}).

By the equality (\ref{disc spec h(p)}) for the study $\sigma_{disc}
(h(p))$ we construct the determinant $det [E+ A_0(p\,; z)].$

 Let the number $z \in {\bf C}\setminus \sigma_{ess}
(h(p))$ be an eigenvalue of the operator $h(p),\,p \in {\bf T}^3$
and $f=(f_0,f_1)\in {\cal H}^{(0,1)}$ be the corresponding
eigenvector, i.e. the equation $h(p)f=zf$ or the system of equations
$$
(w_1(p)-z)f_0+ \frac{1}{\sqrt{2}} \int\limits_{{\bf T}^3}
v_1(s)f_1(s)ds=0;
$$
\begin{equation}\label{sistem h_3 example}
 \frac{1}{\sqrt{2}} v_1(q)f_0+(w_2(p,q)-z)f_1(q)-\int\limits_{{\bf T}^3}
 \sum_{i=1}^3 \cos (s_i - q_i) f_1(s)ds=0
\end{equation}
has a nontrivial solution $f=(f_0,f_1)\in {\cal H}^{(0,1)}.$

Denote
\begin{equation}\label{d_i}
 d_i=\int\limits_{{\bf T}^3}
 \cos s_i \, f_1(s)ds,\quad e_i=\int\limits_{{\bf T}^3}
 \sin s_i \, f_1(s)ds.
\end{equation}

Since $z \in {\bf C}\setminus \sigma_{ess} (h(p))$ from the second
equation of system (\ref{sistem h_3 example}) we  find
\begin{equation}\label{f_1 example}
f_1(q)=\frac{\sum\limits_{i=1}^3(d_i \cos q_i+e_i \sin
q_i)}{w_2(p,q)-z}- \frac{1}{\sqrt{2}} \frac{v_1(q) f_0}{w_2(p,q)-z}.
\end{equation}

For any $p\in {\bf T}^3$ we define the following continuous
functions in ${\bf C}\setminus [m(p), M(p)]$ by
$$
a_{ij}(p\,; z)=-\int\limits_{{\bf T}^3} \frac{\cos s_i \, \cos s_j
\, ds} {w_2(p,s)-z}\,; i,j=1,2,3;
$$
$$
b_{i}(p\,; z)=\int\limits_{{\bf T}^3} \frac{\sin^2 s_i \, ds}
{w_2(p,s)-z},\quad c_{i}(p\,; z)= \frac{1}{\sqrt{2}}
\int\limits_{{\bf T}^3} \frac{\cos s_i \, v_1(s) ds} {w_2(p,s)-z}\,;
i=1,2,3;
$$
$$
D_0(p\,; z)=w_1(p)-z-\frac{1}{2} \int\limits_{{\bf T}^3}
\frac{v_1^2(s)ds}{w_2(p,s)-z};
$$
$$
D_i(p\,; z)=1-a_{ii}(p\,; z),\quad \Delta_{i}(p\,; z)=1-b_{i}(p\,;
z), \quad i=1,2,3,
$$
$$
\Delta_{4}(p\,; z)=\left \vert \begin{array}{llll}
D_0(p\,; z) & c_{1}(p\,; z) & c_{2}(p\,; z) & c_{3}(p\,; z)\\
c_{1}(p\,; z) & D_1(p\,; z) & a_{12}(p\,; z) & a_{13}(p\,; z) \\
c_{2}(p\,; z) & a_{21}(p\,; z) & D_2(p\,; z) & a_{23}(p\,; z)\\
c_{3}(p\,; z) & a_{31}(p\,; z) & a_{32}(p\,; z) & D_3(p\,; z)
\end{array}\right \vert.
$$

Substituting the expression (\ref{f_1 example}) for $f_1$ into the
first equation of system (\ref{sistem h_3}) and the equalities
(\ref{d_i}), we get that the equality
$$
det[E+ A_0(p\,; z)]=\prod\limits_{i=1}^4 \Delta_{i}(p\,; z).
$$

Therefore, by the equality (\ref{disc spec h(p)}) we obtain that
$$
\sigma_{disc}(h(p))=\{z\in {\bf C}\setminus \sigma_{ess}(h(p)):
\prod\limits_{i=1}^4 \Delta_{i}(p\,; z)=0\},\,p\in {\bf T}^\nu.
$$

By Lemma \ref{spektr of channel} and Theorem \ref{ess of H} we have
that
$$
\sigma_{ess}(H)=\bigcup\limits_{p\in {\bf T}^3} \sigma_{disc}(h(p))
\cup [m; M].
$$

Note that if the function $w_2(\cdot,\cdot)$ has form (\ref{example
for w_2}), then we have that $[m; M]=[0; 12].$

\vspace{0.2cm}

{\bf ACKNOWLEDGEMENTS.} The authors are grateful to Prof. Dr. H.
Spohn for stimulating discussions on the results of the paper. This
paper was completed during the visit of the second author at the
Abdus Salam International Centre for Theoretical Physics (ICTP),
Trieste, Italy. He would like to thank to ICTP for the kind
hospitality and support, and the Commission on Development and
Exchanges of the International Mathematical Union for the travel
grant.

\vspace{0.5cm}

{\sc Mukhiddin I. Muminov\\
Samarkand State University,\\
Faculty of Physics and Mathematics,\\
Department of Mathematical Modeling,\\
15 University Boulevard,\\
Samarkand, 140104, Uzbekistan\\
E-mail: mmuminov@mail.ru}

\vspace{0.2cm}

{\sc Tulkin H. Rasulov\\
Bukhara State University,\\
Faculty of Physics and Mathematics,\\
Department of Algebra and Analysis,\\
11 M. Ikbol street,\\
Bukhara, 200100, Uzbekistan\\
E-mail: rth@mail.ru}

\end{document}